\documentclass{article}
\usepackage{tikz, pgfplots, wrapfig}
\usepackage{tikz-3dplot} 
\usepackage{caption}
\usepackage{color}
\usepackage{subcaption}
\usepackage{amssymb}
\usepackage{amsfonts}
\usepackage{amsmath}
\usepackage{fullpage}
\usepackage{graphicx}
\usepackage[utf8]{inputenc}
\usepackage{amsthm}
\usepackage{textcomp}
\usepackage{flexisym}
\usepackage{float}

\usepackage{xfrac}

\newcommand{\rP}{\mathrm{P\space }} %probability, when spacing isn't a worry
\newcommand{\rE}{\mathrm{E\space }} %expectation, when spacing isn't a worry
\newcommand{\rV}{\mathrm{Var\hspace{0.2mm}}} %expectation, when spacing isn't a worry

\usepackage{authblk}

\title{Counting unique molecular identifiers in sequencing using a multitype branching process with immigration
}
\author[a]{Serik Sagitov} \author[b,c,d]{Anders St\aa hlberg}
\affil[a]{Mathematical Sciences, Chalmers University of Technology and University of Gothenburg, \textit{serik@chalmers.se}} 
\affil[b]{Sahlgrenska Center for Cancer Research, Department of Laboratory Medicine,
Institute of Biomedicine, University of Gothenburg,  \textit{anders.stahlberg@gu.se}}
\affil[c]{Wallenberg Centre for Molecular and Translational Medicine,  University of Gothenburg}
\affil[d]{Region Västra Götaland, Sahlgrenska University Hospital, Department of Clinical Genetics and Genomics, Gothenburg, Sweden}

\begin{document}
\maketitle

\begin{abstract}
Detection of extremely rare variant alleles, such as tumour DNA, within a complex mixture of DNA molecules is experimentally challenging due to sequencing errors. Barcoding of target DNA molecules in library construction for next-generation sequencing provides a way to identify and bioinformatically remove polymerase induced errors. During the barcoding procedure involving $t$ consecutive PCR cycles, the DNA molecules become barcoded by unique molecular identifiers (UMI). Different library construction protocols utilise different values of $t$. The effect of a larger $t$ and imperfect PCR amplifications is poorly described. 

This paper proposes a branching process with growing immigration as a model describing the random outcome of $t$  cycles of PCR  barcoding. Our model discriminates between five different amplification rates $r_1$, $r_2$, $r_3$, $r_4$, $r$ for different types of molecules associated with the PCR barcoding procedure. We study this model by focussing on $C_t$, the number  of clusters of molecules sharing the same 
UMI, as well as  $C_t(m)$, the number of UMI clusters of size $m$. Our main finding is a remarkable asymptotic pattern valid for moderately large $t$. It turns out that 
$\rE(C_t(m))/\rE(C_t)\approx 2^{-m}$ for $m=1,2,\ldots$, regardless of the underlying parameters $(r_1,r_2,r_3,r_4,r)$. The knowledge of the quantities $C_t$ and $C_t(m)$ as functions of the experimental parameters $t$ and $(r_1,r_2,r_3,r_4,r)$ will help the users to draw more adequate conclusions from the outcomes of different sequencing protocols.

\end{abstract}
{\bf Keywords:} 
 PCR barcoding, PCR amplification rate, UMI cluster, tree-bookkeeping, PCR branching process, decomposable multitype Galton-Watson process, growing immigration.

\section{Introduction}

\begin{figure}[h!]
\centering
\includegraphics[width=7cm]{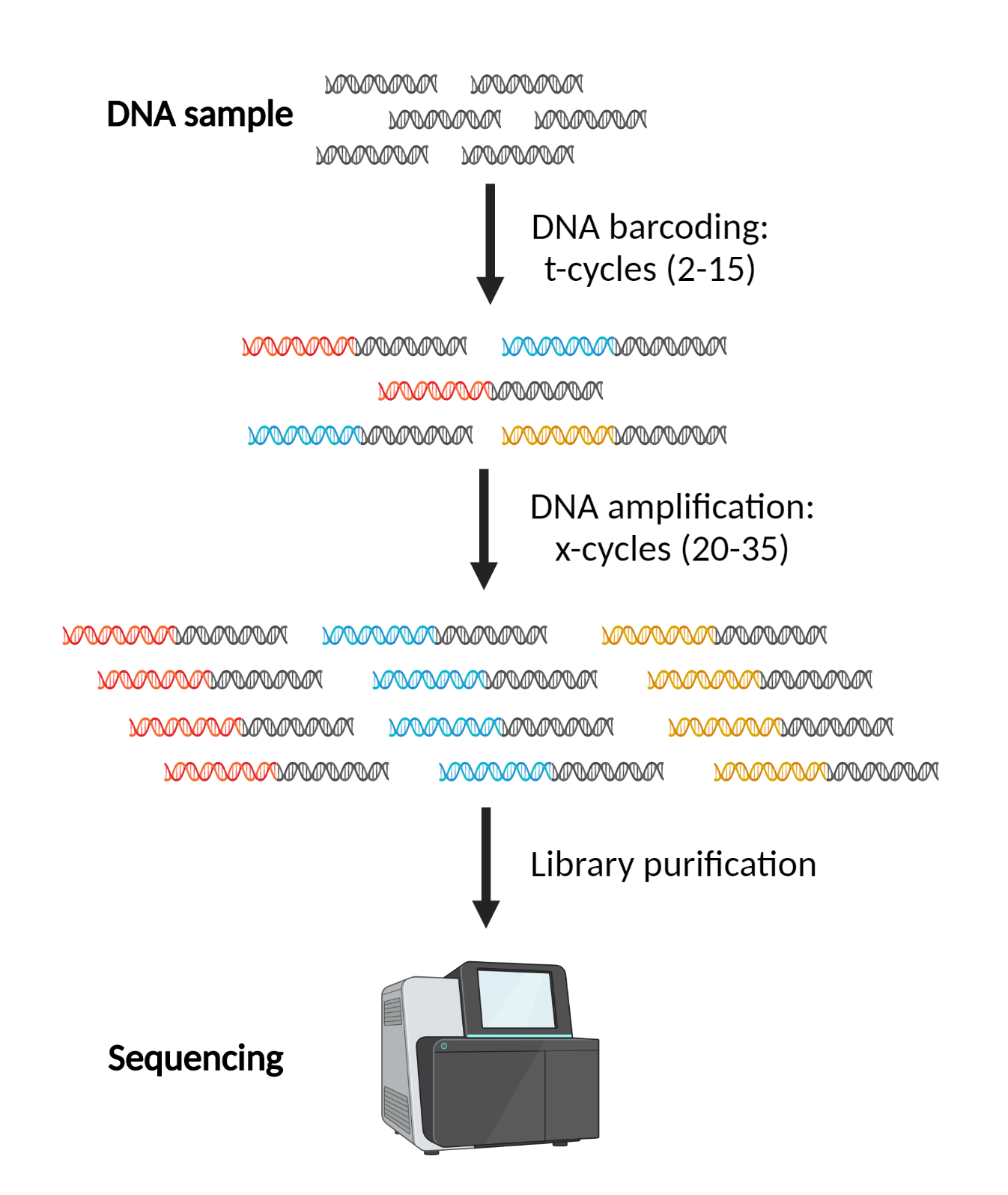}
 \caption{Overview of an ultrasensitive sequencing protocol  (the figure is created using BioRender.com).  The experimental workflow consists of two PCR steps. In the first, barcoding PCR step, the UMIs are attached to target DNA molecules during $t$ cycles of amplification. In the second, adapter PCR step, sequencing adapters are attached to the barcoded DNA sequences during $x$ cycles of amplification. Typical numbers for the parameters $t$ and $x$ are shown. Finally, libraries are purified and sequenced.
  }
 \label{f02}
\end{figure}

Massive parallel sequencing is implemented in a wide range of applications within basic research and for clinical practice. Numerous protocols and technologies are developed to accurately detect and quantify differences in molecular sequences and detect variants. In cancer management, sequencing is applied in diagnostics to identify mutations that can be targeted with specific therapies. The standard massive parallel sequencing techniques can detect variants with frequencies down to the range of 1–5\% \cite{Ste,Xu}. However, this sensitivity is not sufficient for several emerging applications. For example, detection of circulating  tumor-DNA in liquid biopsies requires technologies that have the ability to detect variants with frequencies lower than $0.1$\% in clinically relevant samples \cite{And,Ign,Hei}. The main source of sequencing noise is due to the polymerase induced errors that occur during library construction and sequencing \cite{Fil}. 

To reduce the sequencing noise,   unique molecular identifiers (UMIs), also known as DNA barcodes, can be used to enable ultrasensitive sequencing  \cite{Kin}. The UMIs typically consist of 8-12 randomised nucleotides that are experimentally attached to each target DNA molecule.
The UMIs are introduced through a limited number of PCR cycles followed by a general amplification step (Figure \ref{f02}). After sequencing, all reads with the same UMI can be tracked back to the original DNA molecule, allowing to control the polymerase induced errors and quantification biases.

Experimentally it is challenging to introduce UMIs, since randomised sequences easily produce non-specific PCR products. To address this challenge, several barcoding PCR cycles can be applied, which simplifies the experimental protocol  \cite{Coh,AS}. However, the number of different UMIs and their distribution is not easily estimated with the increasing number of barcoding PCR cycles, limiting the use of UMIs in different applications.

\begin{figure}[H]
\centering
\includegraphics[width=14cm]{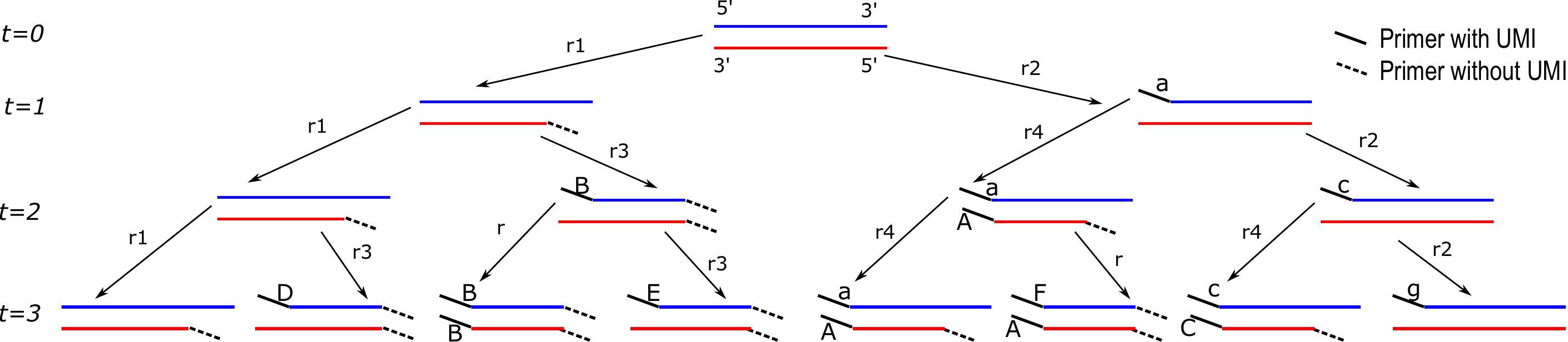}
 \caption{%Schematic representation of the output of the first three cycles for the barcoding PCR step with perfect amplification. The two target DNA strands are shown in blue and red (sense and antisense, respectively). The forward primers with the UMIs and the reverse primers are shown as the solid and dashed segments at an angle.  %A left arrow indicates the amplification of the sense strand generating a new antisense strand with the same barcode inherited (if any) from the parental molecule. A right arrow indicates the amplification of the antisense strand resulting in a novel barcode emerging on the new sense strand.  The capital letters mark different barcodes at the complete molecules, having both forward and backward primers. The solid arrows mark the successful PCR amplification, these arrows are labeled by four possibly  different amplification rates: $r_1,r_2,r_3,r_4$. \\
 Schematic representation for three cycles of barcoding PCR assuming that all fourteen PCR amplifications were successful. The two target DNA strands are shown in blue and red (sense and antisense, respectively). The forward primers with UMI and the reverse primers are shown as the solid and dashed black segments. A single-stranded molecule needs primer sequences at both ends to be complete for the use in downstream PCR. Capital letters mark complete molecules with different UMIs, while non-capitalized letters indicate incomplete molecules with different UMIs.  The amplification rates $r_1,r_2,r_3,r_4,r$, in general, may differ from each other.}
 \label{f00}
\end{figure}

Figure  \ref{f00}  depicts the outcome of three perfectly successful barcoding PCR cycles. For each double-stranded molecule, the upper segment represents a single-stranded molecule in the direction $5\textprime\to3\textprime$ and the lower segment represents a single-stranded molecule in the direction $3\textprime\to5\textprime$. 
The target double-stranded DNA molecule is placed at the top level, $t=0$. 
According to Figure  \ref{f00}, the first PCR cycle produces two double-stranded molecules shown at the level $t=1$:  the left pair, consisting of the target sense molecule plus the antisense molecule with a reverse primer, and the right pair, consisting of the target antisense molecule plus the sense molecule with a forward UMI primer.
 All four single stranded molecules at $t=1$ are incomplete with one generated UMI. %(besides the target molecules, the incomplete molecules are those that miss either the forward or reverse primer), with only one barcode  $a$ generated.  
 The complete molecules %, containing both a barcode and a reverse primer, 
start appearing at the level $t=2$. With perfect PCR amplifications, the sizes of UMI clusters grow geometrically as shown by Table 1. For example at $t=3$, as illustrated on Figure  \ref{f00}, there are six UMI clusters labelled by A, B, C, D, E, F, with  the clusters B and D having size two, and the clusters A, C, E, F being singletons.

\begin{table}[htp]
\caption{The UMI cluster numbers for the perfect PCR amplifications.  
}
\begin{center}
\begin{tabular}{l|ccccc|cc}
Cluster size&1&2&3&4&5&Total 
\\\hline
$t=2$ &2&0&0&0&0&2\\ 
$t=3$ &4&2&0&0&0&6\\ 
$t=4$ &8&4&2&0&0&14\\ 
$t=5$ &16&8&4&2&0&30\\ 
$t=6$ &32&16&8&4&2&62
\end{tabular}
%\qquad \qquad 
%\begin{tabular}{l|ccccc|cc}
%Cluster size&1&2&3&4&5&Total 
%\\\hline
%$t=2$ &0&0&0&0&0&0\\ 
%$t=3$ &1&0&0&0&0&1\\ 
%$t=4$ &4&0&0&0&0&4\\ 
%$t=5$ &4&2&0&0&0&6\\ 
%$t=6$ &6&1&2&0&0&9
%\end{tabular}

\end{center}
\label{default}
\end{table}%

%\begin{figure}[h!]
%\centering
%\includegraphics[width=16cm]{Tre01.pdf}
% \caption{Schematic representation of the first three cycles of the barcoding PCR step. The two DNA strands are shown in blue and red (sense and antisense, respectively). The forward primers with the Unique Molecular Identifiers (UMIs) and the reverse primers are shown as the solid and dashed segments at an angle. Numbers indicate the single-stranded DNA molecules without UMIs, while the labels that starts with a letter indicate the single-stranded DNA molecules with UMIs. Different such labels indicate specific UMIs. Labels that include small letters represent the single-stranded molecules with UMIs that will be amplified in the barcode PCR step but not in the adapter PCR step. 
%  }
% \label{f00}
%\end{figure}

\begin{figure}[H]
\centering
\includegraphics[width=15cm]{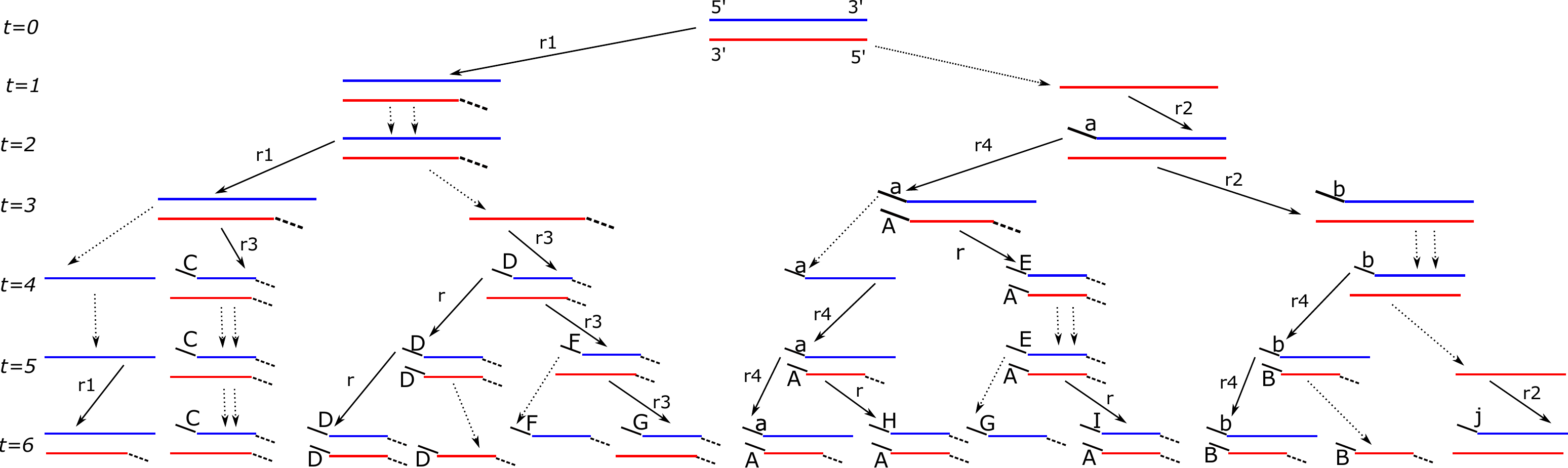}
 \caption{A possible output of the first six cycles for the barcoding PCR step with imperfect amplifications.  The failed PCR amplifications are marked by the dashed arrows and dashed double-arrows. 
 }
 \label{fin}
\end{figure}

In practice, PCR amplifications are imperfect and a single-stranded molecule gets successfully amplified with a probability $r$, which we call the amplification rate. Despite the attempts to optimise the primers and reaction conditions, the PCR amplification rates are
rarely close to 1. With imperfect amplifications, Figure 2 and Table 1 should be modified to reflect the fact that the outcome of each PCR cycle is random.  %and the bookkeeping of the UMI clusters produced by a larger number of PCR cycles is a challenging task. 
Figure  \ref{fin}
illustrates a possible realisation of six barcoding PCR cycles, summarised by Table 2. In particular, at $t=6$, as shown on the bottom rows of  Figure  \ref{fin} and Table 2, A and D form UMI clusters of size 3, B is a UMI cluster of size 2, and the clusters C, E, F, G, H, I are singletons.
\begin{table}[htp]
\caption{The UMI cluster numbers of imperfect PCR cycles depicted in Figure \ref{fin}. }
\begin{center}
\begin{tabular}{l|ccccc|cc}
Cluster size&1&2&3&4&5&Total 
\\\hline
$t=2$ &0&0&0&0&0&0\\ 
$t=3$ &1&0&0&0&0&1\\ 
$t=4$ &4&0&0&0&0&4\\ 
$t=5$ &4&2&0&0&0&6\\ 
$t=6$ &6&1&2&0&0&9
\end{tabular}
\end{center}
\label{default}
\end{table}%

The PCR amplification rate is dependent on both sequence context and sample
quality. In particular, the target DNA molecules are often long, containing thousands
or even millions base pairs. Moreover, some molecules may have inhibitors attached to the DNA.  As a rule, the amplification rates of the original molecules are smaller  compared to the later formed DNA molecules. Addressing these features, our model discriminates between five amplification rates $(r_1,r_2,r_3,r_4,r)$ as indicated by Figures  \ref{f00} and \ref{fin}, assuming that
\begin{equation}\label{Diana}
 0<r_1,r_2\le  r_3,r_4\le r\le1.
\end{equation}
Here, $r_1$ and $r_2$ represent the amplification rates of the original sense and antisense DNA molecules, respectively. The rates $r_1$ and $r_2$ may differ, since their sequence contexts are different, in so far as two complementary sequences have different nucleotide sequences. The parameters $r_3$ and $r_4$ refer to the amplification rates of the DNA molecules that have been amplified once. Again, $r_3$ and $r_4$ may be different since they have different sequence contexts. We assume that both $r_3$ and $r_4$ are larger than each of the rates $r_1$ and $r_2$, since the DNA molecules become shorter after being amplified once. Finally, we assume that the shortest amplicon that is amplified exponentially is amplified with the highest rate $r$ disregarding the difference between the sense and antisense strands.

 This paper introduces  and studies a mathematical model for the outcome of the barcoding PCR experiment starting from a single double-stranded DNA molecule. 
Our stochastic model is defined in terms of the five parameters $(r_1,r_2,r_3,r_4,r)$ using the framework of the multitype Galton-Watson processes \cite{Hac}.
The use of branching processes as a stochastic model for counting the molecules in the repeated PCR amplification cycles is well established in the literature, see \cite{Hac, JK, Kim, Kra, Lal} and references therein. In \cite{Pfl}  the branching process approach is applied to the second, adapter PCR step mentioned in Figure \ref{f02}. However, to our knowledge, the use of branching processes is novel for modelling the barcoding PCR step. The multitype branching process with immigration of this paper is a special example of the multitype Galton-Watson processes with neutral mutations examined in \cite{Be}. %It is an interesting mathematical object whose properties beyond the first moments will be studied in forthcoming papers. 

\section{Results}\label{Res}

 We study the possible outcomes of the barcoding PCR step, starting from a single double-stranded DNA molecule.
Our stochastic model for counting the unique UMIs is built upon an efficient bookkeeping system for the DNA barcoding procedure, presented in Section \ref{S0}.  We discriminate between six different types of single-stranded molecules emerging during the barcoding PCR procedure and introduce a multitype Galton-Watson process \cite{Hac} with immigration describing the random process of reproduction of the single-stranded molecules, see Section \ref{Sli}. 
 
We study this model by focussing on $C_t$, the number  of clusters of molecules sharing the same 
UMI, as well as  $C_t(m)$, the number of UMI clusters of size $m$. The underlying branching properties of the model yield recursive formulas for the expected values $\rE(C_t)$ and $\rE(C_t(m))$, see Section \ref{Sexp}. Our main finding, based on the analysis of the proposed multitype Galton-Watson process, is the following asymptotic result
\begin{equation}\label{Zel}
\rE(C_t(m))/\rE(C_t)\to 2^{-m},\quad m\ge 1,\quad t\to\infty.
\end{equation}
The remarkable feature of this relationship is that the limits are the same irrespectively of the parameter values $(r_1,r_2,r_3,r_4,r)$.  This approximation may work well already at $t=10$, as illustrated by Figure \ref{f5} below.

\subsection{Tree-bookkeeping system for barcoding PCR  %of unique molecular identifiers
}\label{S0}

\begin{figure}[H]
\centering
\includegraphics[width=15cm]{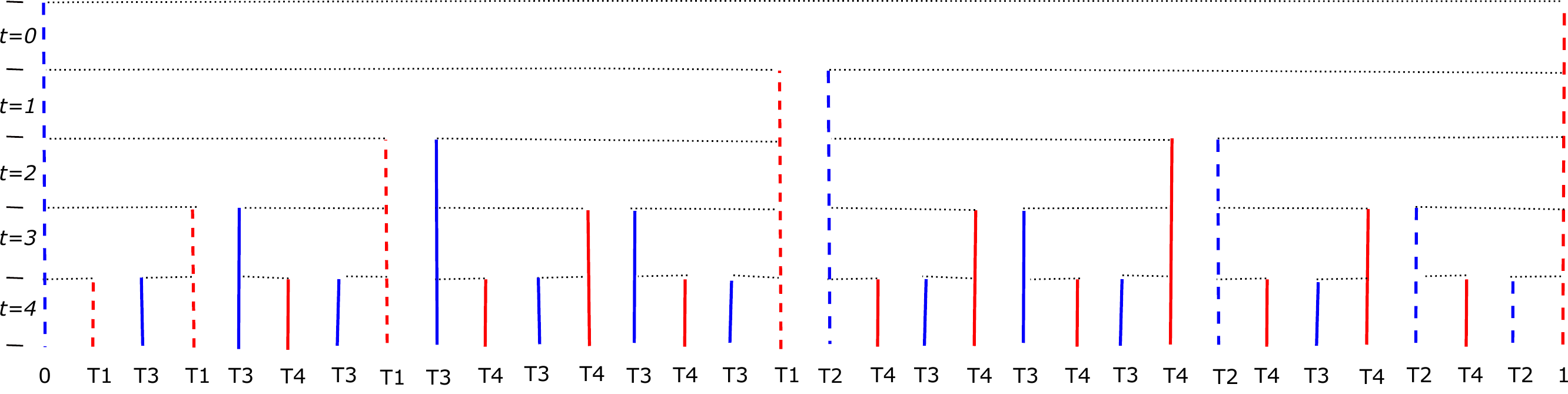}
 \caption{A tree-graph summary of the four cycles of perfect barcoding PCR.
 Each vertical lineage represents a molecule that, once appeared, persists over the succeeding PCR cycles. The dashed and solid branches discriminate between the incomplete and complete molecules. Labels $0, 1, T_1, T_2, T_3 ,T_4$ indicate different types of the  molecules with and without UMIs, where the different levels $t$ of the tree represent the consecutive PCR cycle numbers.
 }
 \label{f2}
\end{figure}

In this section we introduce a convenient bookkeeping system for the outcome of multiple barcoding PCR cycles. We start by considering the simple case of perfect PCR amplifications. Compared to the schematic representation of Figure \ref{f00}, the tree-graph approach of Figure  \ref{f2} allows us to neatly depict more than three PCR cycles using the same space.  At any tree level $t$, the vertical branches (lineages) of the tree on  Figure  \ref{f2} are  pairwise connected designating the double-stranded molecules of Figure  \ref{f00}. 
%Using this bookkeeping system we study the population of clusters of sizes $m=1,\ldots$ evolving in time described by the parameter $t\ge1$. 

Out bookkeeping system distinguishes between six different types of the  tree lineages. At any level $t$ there is exactly one target sense lineage labeled by 0, and one  target antisense lineage labeled by 1. The other four type of lineages $T_1,T_2,T_3,T_4$ are defined by the following lineage generation rules:
\begin{align}
&0\to T_1,\quad 1\to T_2,\quad T_1\to T_3,\quad T_2\to T_4,\quad T_3\to T_4, \quad T_4\to T_3, \label{Ti}
\end{align}
where
\begin{description}
\item[ \quad -]  the lineages $0,1,T_1, T_2$ represent incomplete molecules, 
\item[ \quad -]  the lineages $T_3, T_4$ represent complete molecules,
\item[ \quad -] newly generated lineages $T_2$ and $T_3$ represent molecules with a novel UMI,
\item[ \quad -]  the lineages $T_4$ represent molecules that inherit the UMI of the parental molecule.
\end{description}

Let $Z_t^{(i)}$ stand for the number of the $T_i$-lineages at the level $t$, for $i=1,2,3,4$ and $t\ge0$.
Given 
\begin{align}\label{one}
(r_1,r_2,r_3,r_4,r)=(1,1,1,1,1),
\end{align}
 the evolution of the vector $(Z_t^{(1)},Z_t^{(2)},Z_t^{(3)},Z_t^{(4)})$  is deterministic and in accordance with rule \eqref{Ti} satisfies the following recursions
\begin{align*}
 Z_t^{(1)}&=Z_{t-1}^{(1)}+1,\\ 
 Z_t^{(2)}&=Z_{t-1}^{(2)}+1,\\
 Z_t^{(3)}&=Z_{t-1}^{(3)}+Z_{t-1}^{(1)}+Z_{t-1}^{(4)},\\
 Z_t^{(4)}&=Z_{t-1}^{(4)}+Z_{t-1}^{(2)}+Z_{t-1}^{(3)},
\end{align*}
valid for $t\ge1$ under the initial condition 
\begin{align}\label{init}
Z_0^{(1)}=Z_0^{(2)}=Z_0^{(3)}=Z_0^{(4)}=0.
\end{align}
From these recursions it is easy to see that 
$Z_t^{(1)}=Z_t^{(2)}=t-1$,
implying $Z_t^{(3)}=Z_{t}^{(4)}$. Moreover,
\[ Z_t^{(3)}=2Z_{t-1}^{(3)}+t-2,\]
so that
\[ Z_t^{(2)}+Z_t^{(3)}=2(Z_{t-1}^{(2)}+Z_{t-1}^{(3)})+1,\]
yielding
\begin{equation}\label{z23}
Z_t^{(2)}+Z_t^{(3)}=2^{t}-1,\quad t\ge 0.
\end{equation}

At any given level $t$, we split the set of  complete lineages into the clusters of lineages representing the UMI clusters of molecules sharing the same UMI. According to our bookkeeping system, there are two different types of lineage clusters: \textit{$T_2$-clusters} 
stemming from the $T_2$-lineages and \textit{$T_3$-clusters} stemming from the $T_3$-lineages. A $T_2$-cluster consists of the $T_4$-lineages, which are the daughter lineages of the stem $T_2$-lineage, with the stem $T_2$-lineage not being  part of the cluster as it represents an incomplete molecule. A $T_3$-cluster consists of the stem $T_3$-lineage and its daughter $T_4$-lineages.

If $C_t(m)$ is the number of $T_2$ and $T_3$-clusters of size $m$ observed at the level $t$ of the lineage tree, then
\begin{equation}\label{zt}
C_t=C_t(1)+\ldots+C_t(t-1)
\end{equation}
gives the total number of lineage clusters at the level $t$. Observe that in the special case \eqref{one}, we have
\[C_t=Z_t^{(2)}+Z_t^{(3)}-1,\]
so that by \eqref{z23},
\begin{equation}\label{jul}
C_t=2^t-2, \quad  t\ge1.
\end{equation}
Furthermore, under \eqref{one},  we get
\begin{equation}\label{elin}
C_{t+1}(1)= C_t+2, \quad C_{t+1}(m+1)=C_t(m), \quad m\ge2, \quad t\ge1.
\end{equation}
The first equality in \eqref{elin} says that each cluster at the level $t$  produces  at the next level $t+1$ one novel $T_3$-cluster, in addition to a new $T_2$-singleton and a new $T_3$-cluster generated by a $T_1$-lineage.  (By a $T_2$-singleton we mean a $T_2$-lineage that has not yet produced a daughter lineage.) 
The second equality in \eqref{elin} says that each cluster of size $m$ at the level $t$ turns into a cluster of size $m+1$ at the next level $t+1$.
By \eqref{jul} and \eqref{elin}, 
\begin{equation}\label{selin}
C_{t}(m)= 2^{t-m}, \quad m\ge1,\quad t\ge m+1,
\end{equation}
cf Table 1. Relations  \eqref{jul} and  \eqref{selin} imply an interesting rule of thumb saying that given \eqref{one}, at any given $t$, the increase of the cluster size by 1 reduces the number of clusters by half. More precisely,
\begin{equation}\label{zel}
C_{t}(m)/C_t\to 2^{-m},\quad m\ge 1,\quad t\to\infty.
\end{equation}
\begin{figure}
\centering
\includegraphics[width=15cm]{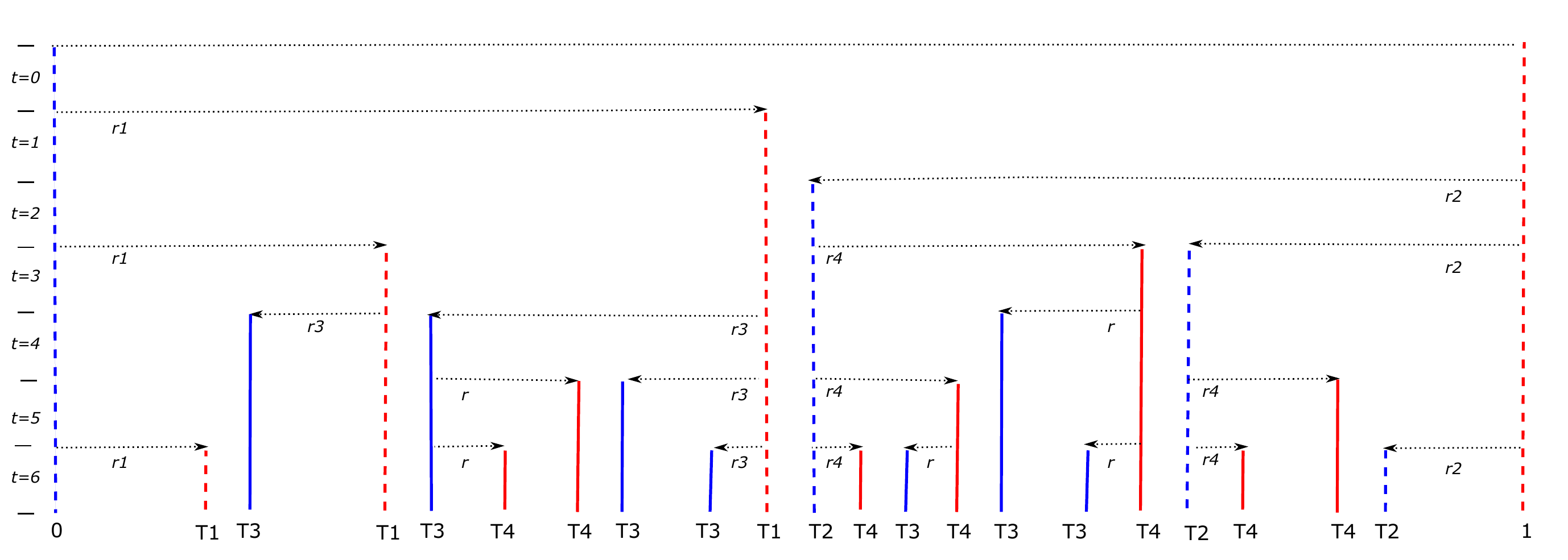}
 \caption{The tree view of the Figure \ref{fin}. Depending on the corresponding type of the underlying DNA molecule, the five amplification rates $r_1,r_2,r_3,r_4,r$, are assigned to the corresponding horizontal arrows.}
 \label{f3}
\end{figure}

If \eqref{one} does not hold, so that some PCR amplifications may fail, the cluster numbers $C_t(m)$ and $C_t$ become random. In the next section, we obtain recursion relations for the corresponding mean values,
from which we will derive \eqref{Zel}, a far-reaching extension of the deterministic relation \eqref{zel}.

%\section{Average cluster size}\label{Se}
%Choose a lineage at random, what its cluster size? Distribution of the cluster size
%\[\tilde p_{t}(m)=\frac{mp_{t}(m)}{\sum_{i=1}^{t-1}ip_t(i)}\]

\subsection{Multitype branching process with immigration }\label{Sli}

We are going to write $ \xi\sim\text{Ber}(r)$ to say that the random variable $ \xi$ has a Bernoulli distribution with 
$$\rP(\xi=1)=r,\quad \rP(\xi=0)=1-r.$$
Assuming that the amplification rates $(r_1,r_2,r_3,r_4,r)$ satisfy \eqref{Diana}, we restate \eqref{Ti} in the form
\begin{align}
&0\to 0+\xi_0T_1,\quad T_1\to T_1+\xi^{(1)}T_3,  \quad T_3\to T_3+\xi^{(3)}T_4,\label{Tber}\\
&1\to 1+\xi_1T_2,\quad T_2\to T_2+\xi^{(2)}T_4,   \quad  T_4\to T_4+\xi^{(4)}T_3,  \label{Tbe}
\end{align}
involving six random inputs
\[\xi_0\sim\text{Ber}(r_1),\quad \xi_1\sim\text{Ber}(r_2),\quad \xi^{(1)}\sim\text{Ber}(r_3),\quad \xi^{(2)}\sim\text{Ber}(r_4),\quad \xi^{(3)},\xi^{(4)}\sim\text{Ber}(r).  \]
We clarify these relations by referring to $T_3\to T_3+\xi^{(3)}T_4$, which says that a $T_3$-lineage existing at any given level $t$, at the next level $t+1$ infallibly reproduces itself and produces a new lineage of type $T_4$ with probability $r$.

The sequence of random vectors 
\begin{align} \label{Stieg}
\{(Z_t^{(1)},Z_t^{(2)},Z_t^{(3)},Z_t^{(4)}):\quad  t=0,1,\ldots\}
\end{align}
forms a Markov chain with the initial state \eqref{init}. Treating relations \eqref{Tber} and  \eqref{Tbe} as the reproduction rules involving four types of individuals along the dicrete time $t$, we may view this Markov chain as a multitype branching process with immigration \cite{Mod}. 
This is an example of a decomposable multitype branching process, since the types $T_3$ and $T_4$ may generate each other but not the types $T_1$ and $T_2$. There are two sources of immigration for this four-type branching process: the $0$-lineage  generates the  $T_1$-individuals at the rate $r_1$, and the $1$-lineage generates the $T_2$-individuals at the rate $r_2$. The types $T_1$ and $T_2$ without directly communicating with each other, give rise to the types $T_3$ and $T_4$ respectively. %The types $T_3$ and $T_4$ do reproduce each other, but not the types $T_1$ and $T_2$.

The reproduction rules \eqref{Tber} and  \eqref{Tbe} yield the following recursive relations
\begin{align}
 Z_t^{(1)}&=Z_{t-1}^{(1)}+\xi_{t,0},\quad
Z_t^{(2)}=Z_{t-1}^{(2)}+\xi_{t,1},  \label{Odes}\\
Z_t^{(3)}&=Z_{t-1}^{(3)}+\sum_{j=1}^{Z_{t-1}^{(1)}}\xi_{t,j}^{(1)}+\sum_{j=1}^{Z_{t-1}^{(4)}}\xi_{t,j}^{(4)},  
\qquad
Z_t^{(4)}=Z_{t-1}^{(4)}+\sum_{j=1}^{Z_{t-1}^{(2)}}\xi_{t,j}^{(2)}+\sum_{j=1}^{Z_{t-1}^{(3)}}\xi_{t,j}^{(3)}, \label{Odess}
\end{align}
involving independent Bernoulli random values
\[\xi_{t,0}\sim\text{Ber}(r_1),\quad \xi_{t,1}\sim\text{Ber}(r_2),\quad \xi_{t,j}^{(1)}\sim\text{Ber}(r_3),\quad \xi_{t,j}^{(2)}\sim\text{Ber}(r_4),\quad \xi_{t,j}^{(3)},\xi_{t,j}^{(4)}\sim\text{Ber}(r),\]
each indicating whether the underlying PCR amplification is successful or not.

In the current setting, the supercritical four-type branching process \eqref{Stieg} could be described in terms of a single type branching process with a growing immigration. The specific reproduction rules \eqref{Tber} and  \eqref{Tbe} allow the types $T_3$ and $T_4$ to be treated as a single type, say $T$, such that the type $T$ individuals produce $(1+r)$ offspring on average. In terms of the process  \eqref{Stieg}, the number $Z_t$ of $T$-individuals at time $t$ can be expressed as the sum
$$Z_t=Z_t^{(3)}+Z_t^{(4)},$$ 
and the Markov chain $(Z_t)_{t\ge0}$ can be treated as a branching process with growing immigration. By \eqref{Odess}, the number of type $T$ immigrants at time $t$ is given by the sum  $I_t=I_t^{(1)}+I_t^{(2)}$  of two independent random variables 
\begin{align*}
I_t^{(1)}=\sum_{j=1}^{Z_{t-1}^{(1)}}\xi_{t,j}^{(1)},\qquad I_t^{(2)}= \sum_{j=1}^{Z_{t-1}^{(2)}}\xi_{t,j}^{(2)}
\end{align*}
 having binomial  distributions $I_t^{(1)}\sim\text{Bin}(t,r_1r_3)$ and $I_t^{(2)}\sim\text{Bin}(t,r_2r_4)$. Observe that the generating function for the number of immigrants $h_t(s)=\rE(s^{I_t})=\rE(s^{I_t^{(1)}})\rE(s^{I_t^{(2)}})$ is computed explicitly
\begin{align}\label{Imm}
h_t(s)=(1-r_1r_3+r_1r_3s)^t(1-r_2r_4+r_2r_4s)^t,\quad 0\le s\le1,\quad t\ge0.
\end{align}

According to Theorem 2b from Section 4 of   \cite{Rah},
the long term population size growth of the supercritical branching process with growing immigration is regulated by its reproduction rate $(1+r)$:
\begin{align} \label{ibra}
(1+r)^{-t}Z_t\to W \text{ in } L_2,\quad t\to\infty.
\end{align}
Here, the limit $W$ is a strictly positive random variable, whose Laplace transform 
\[\rE(e^{-\lambda W})=\prod_{k=1}^\infty h_k(\lambda(1+r)^{-k})\]
 is determined by the five amplification rates $(r_1,r_2,r_3, r_4, r)$ in terms of the generating functions  \eqref{Imm}
 and the limiting Laplace transform $\phi(\lambda)$ for the branching process without immigration satisfying the functional quadratic equation
\[\phi((1+r)\lambda)=(1-r)\phi(\lambda)+r\phi^2(\lambda).\]

The main concern of this paper is not the decomposable multitype branching process \eqref{Stieg} per se, but certain functionals thereof, especially the number of clusters of size $m$,
\[C_t(m)=X_t(m)+Y_t(m),\quad 1\le m\le t-1,\]
where $X_t(m)$ is the number of $T_2$-clusters and $Y_t(m)$ is the number of $T_3$-clusters of size $m$ at the level $t$. By  \eqref{Odess}, we have
 \begin{align}
 Y_t(1)&=\sum_{j=1}^{Z_{t-1}^{(1)}}\xi_{t,j}^{(1)}+\sum_{j=1}^{Z_{t-1}^{(4)}}\xi_{t,j}^{(4)}+\sum_{j\in\mathbb Y_{t-1}(1)}(1-\xi_{t,j}^{(3)}). \label{Yt1}
 \end{align}
 Here, $\mathbb Y_{t}(m)$ is the set of $T_3$-lineages which have exactly $(m-1)$ daughter lineages of the type $T_4$ at the level $t$.
Furthermore, again by \eqref{Odess},  for $1\le m\le t-1$, 
\begin{align}
 X_t(m)&=\sum_{j\in\mathbb X_{t-1}(m-1)}\xi_{t,j}^{(2)}+\sum_{j\in\mathbb X_{t-1}(m)}(1-\xi_{t,j}^{(2)}), \label{Xtm}\\
Y_t(m)&=\sum_{j\in\mathbb Y_{t-1}(m-1)}\xi_{t,j}^{(3)}+\sum_{j\in\mathbb Y_{t-1}(m)}(1-\xi_{t,j}^{(3)}), \label{Ytm}
\end{align}
where $\mathbb X_{t}(m)$ is the set of $T_2$-lineages which have exactly $m$ daughter lineages of the type $T_4$ at the level $t$, provided $m\ge1$, while $\mathbb X_{t}(0)$ is the set of $T_2$-singletons at the level $t$.
 
Let  $X_t(0)$ be the number of $T_2$-singletons  at the level $t$. (To illustrate, the example of Figure  \ref{f3} gives $X_0(0)=X_1(0)=X_5(0)=0$, $X_2(0)=X_3(0)=X_4(0)=X_6(0)=1$.)
Then, due to \eqref{Odes} and \eqref{Odess}, 
\begin{align}\label{X0}
X_t(0)&=\xi_{t,1}+\sum_{j\in\mathbb X_{t-1}(0)}(1-\xi_{t,j}^{(2)}).
 \end{align}
Since
 \[Z_t^{(2)}=X_t(0)+X_t(1)+\ldots+X_t(t-1),\]
the total number of clusters at the level $t$ equals
\begin{align}\label{EGO}
C_t=C_t(1)+\ldots+C_t(t-1)=X_t(1)+Y_t(1)+\ldots+X_t(t-1)+Y_t(t-1)=Z_t^{(2)}-X_t(0)+Z_t^{(3)}.
 \end{align}
In the expression \eqref{EGO} for the total number of clusters $C_t$, the dominating term is $Z_t^{(3)}$, which according to \eqref{ibra} is of order $(1+r)^{t}$. Observe that relation  \eqref{ibra} implies that both the mean number of clusters $\rE(C_t)$ and the standard deviation $\sqrt{\rV(C_t)}$ are growing proportionally to $(1+r)^{t}$ as $t\to\infty$. %This intuition is supported by the forthcoming relations \eqref{Ukr} and \eqref{Var}.

\subsection{The expected values }\label{Sexp}
\begin{figure}
\centering
\includegraphics[width=5.4cm]{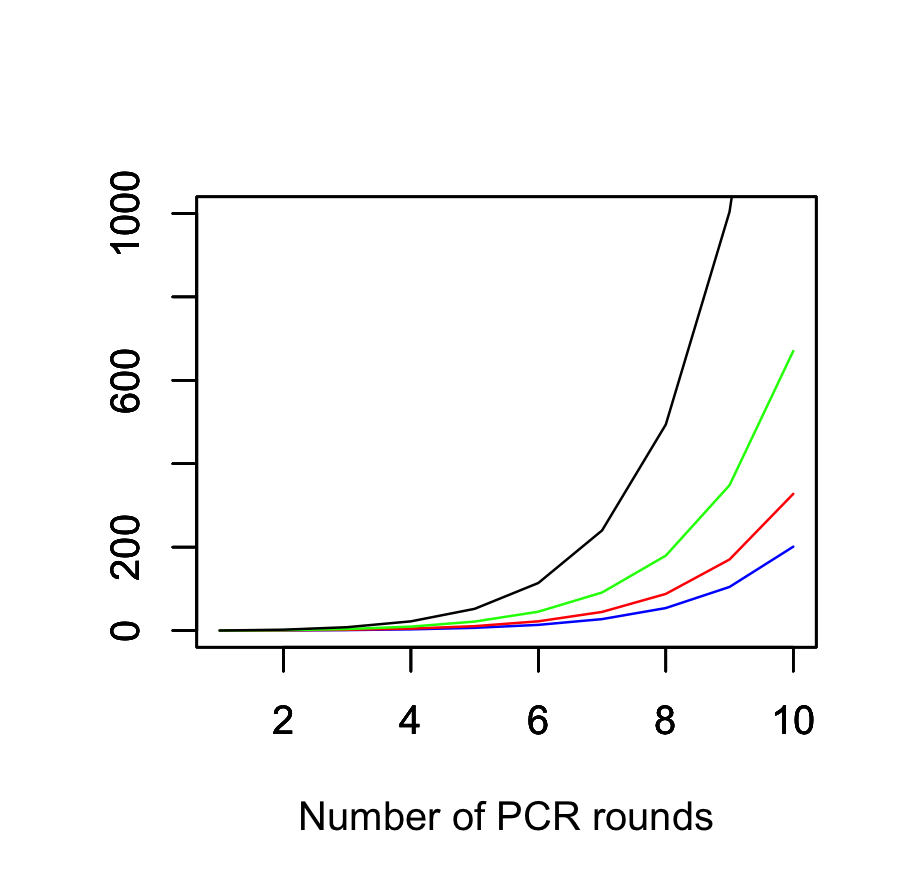}\includegraphics[width=5.4cm]{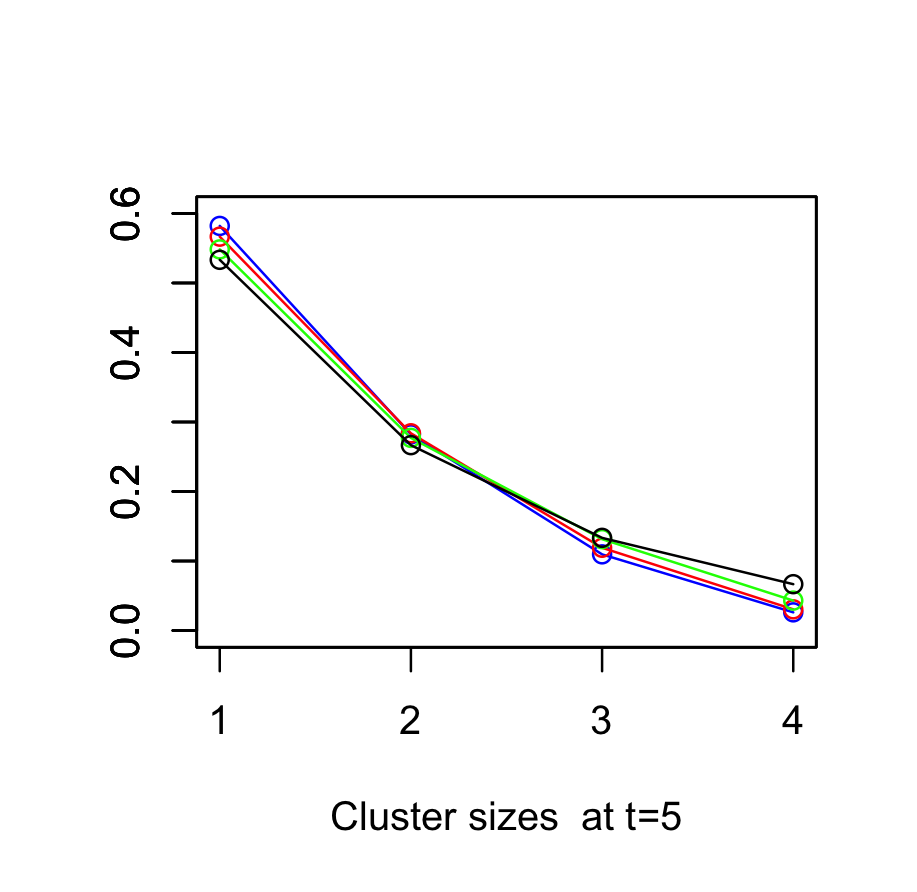} \includegraphics[width=5.4cm]{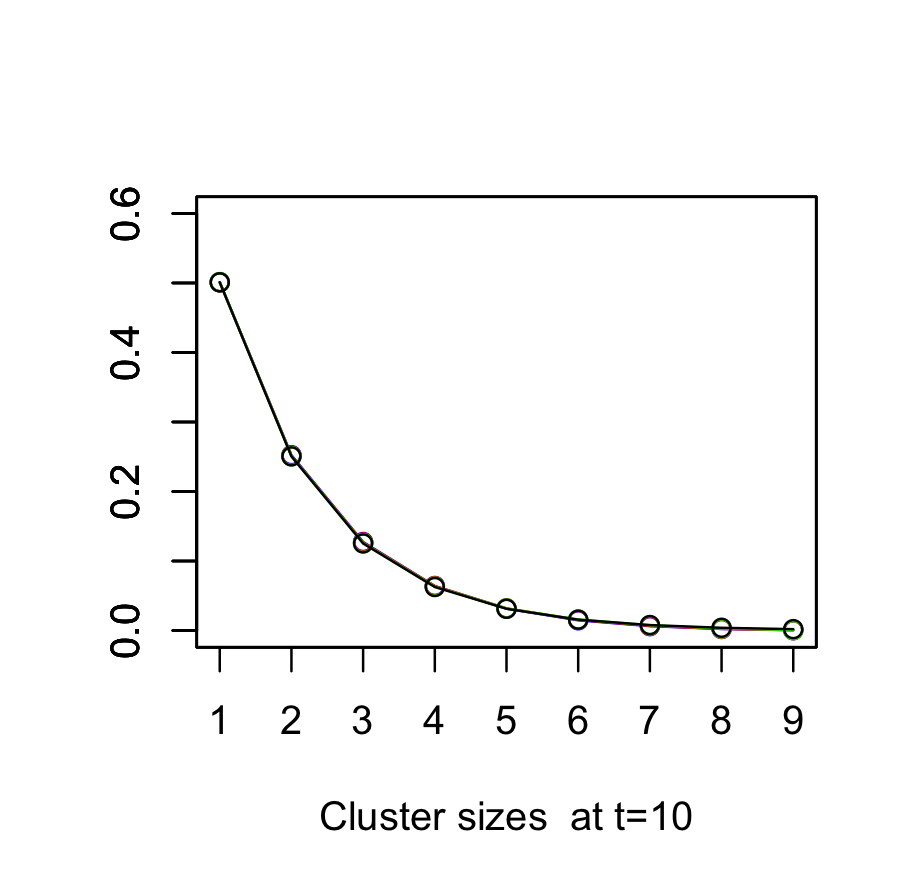}
 \caption{Left panel: the expected values of the total number of clusters $\rE(C_t)$. Middle ($t=5$) and right ($t=10$) panels show the plots of $\rE(C_t(m))/\rE(C_t)$ over the cluster sizes $m=1,\ldots, t-1$. Different colours represent different sets of the parametrs $(r_1,r_2,r_3,r_4,r)$: black $(1,1,1,1,1)$, green $(0.6,0.6,0.7,0.8,0.9)$, red  $(0.4,0.4,0.6,0.7,0.9)$,  and blue $(0.3,0.3,0.4,0.5,0.9)$. }
 \label{f5}
\end{figure}

In this section,  we denote
$$c_t=\rE(C_t),\quad  c_t(m)=\rE(C_t(m)), \quad 1\le m\le t-1.$$
and show first  that
\begin{align} \label{Ukr} 
c_t%&=r_1t+\tfrac{1}{2}r_1(r_2+ r_3)r_4^{-2}(1+r_4)^t-r_1r_3r_4^{-1}t-r_1r_2r_4^{-2}+\tfrac{1}{2}r_1(r_2- r_3)r_4^{-2}(1-r_4)^t-  r_1r_3^{-1}(1-(1-r_3)^{t})\\
%&=\tfrac{1}{2}r_1(r_2+ r_3)r_4^{-2}(1+r_4)^t+r_1(1-r_3r_4^{-1})t-r_1(r_2r_4^{-2}+r_3^{-1})+\tfrac{1}{2}r_1(r_2- r_3)r_4^{-2}(1-r_4)^t+ r_1r_3^{-1}(1-r_3)^{t}\\
&=\alpha (1+r)^t+\alpha_1 t-\alpha_2+\alpha_3(1-r_4)^{t}+\alpha_4(1-r)^t,
\end{align}
where
\[%A_1:=\tfrac{1}{2}r_1(r_2+ r_3)r_4^{-2},
\alpha:=\frac{r_1r_3+r_2r_4}{2r^{2}},\quad \alpha_1:=r_2(1-r_4r^{-1}),\quad \alpha_2:=r_1r_3r^{-2}+r_2r_4^{-1},\quad \alpha_3:=r_2r_4^{-1},\quad %A_5:=\tfrac{1}{2}r_1(r_2- r_3)r_4^{-2}.
\alpha_4:=\frac{r_1r_3-r_2r_4}{2r^{2}},\]
and then derive the main result \eqref{Zel} of this paper. Notice that  $\alpha_4=0$ if $r_1r_3=r_2r_4$, and in the deterministic case \eqref{one}, relation \eqref{Ukr} turns into \eqref{jul}.  Our results concerning the expected values are illustrated by Figure \ref{f5}.  On the right panel of Figure \ref{f5}, the four lines, representing different combinations of the parameter values, almost coincide demonstrating that asymptotic relation \eqref{Zel} works well already for $t=10$.

Put
$$z_t^{(i)}=\rE(Z_t^{(i)}),\quad x_t(m)=\rE( X_t(m)),\quad y_t(m)=\rE( Y_t(m)).$$
The proofs of \eqref{Ukr} and  \eqref{Zel} rely on the recursive relations
\begin{align}
x_t(0)&=r_2+x_{t-1}(0)(1-r_4), \label{xt}\\
y_t(1)&=r_3z_{t-1}^{(1)}+rz_{t-1}^{(4)}+y_{t-1}(1)(1-r),  \label{yt0}\\
 x_t(m)&=x_{t-1}(m-1)r_4+x_{t-1}(m)(1-r_4), \quad 0\le m\le t-1,\label{xtm}\\
y_t(m)&=y_{t-1}(m-1)r+y_{t-1}(m)(1-r), \quad 0\le m\le t-1, \label{ytm}
\end{align}
following from \eqref{X0}, \eqref{Xtm},  \eqref{Yt1},  and \eqref{Ytm}.

\vspace{5mm}
\noindent{\sc Proof of \eqref{Ukr}}. From \eqref{Odes} we obtain 
\[ z_t^{(1)}=r_1+z_{t-1}^{(1)},\quad z_t^{(2)}=r_2+z_{t-1}^{(1)},\quad z_0^{(1)}=z_0^{(2)}=0,\]
so that $z_t^{(1)}=r_1t$, $z_t^{(2)}=r_2t$, and therefore by \eqref{Odess},
\begin{align*}
z_t^{(3)}&=z_{t-1}^{(3)}+r_1r_3(t-1)+ rz_{t-1}^{(4)},\\
z_t^{(4)}&=z_{t-1}^{(4)}+r_2r_4(t-1)+rz_{t-1}^{(3)}.
\end{align*}
This yields
\begin{align*}
z_t^{(3)}+z_t^{(4)}&=(r_1r_3+r_2r_4)(t-1)+(1+r)(z_{t-1}^{(3)}+ z_{t-1}^{(4)}) %\qquad \qquad  \qquad \qquad  \Bigg[ \alpha_t:=r_1(r_2+ r_3)t\Bigg]
\\
&=(r_1r_3+r_2r_4)\sum_{j=1}^{t-1}(t-j)(1+r)^{j-1}%=r_1(r_2+ r_3)r_4^{-2}((1+r_4)^t-r_4t-1)
=2\alpha((1+r)^t-rt-1),
\end{align*}
%\[\sum_{j=1}^{t-1}js^{j-1}=\tfrac{d}{ds}\sum_{j=0}^{t-1}s^{j}=\tfrac{d}{ds}(\tfrac{s^t-1}{s-1})=\frac{ts^{t-1}(s-1)+1-s^t}{(s-1)^2}=(s-1)^{-1}ts^{t-1}-(s-1)^{-2}(s^t-1)\]
where we used the relation
\[\sum_{j=1}^{t-1}(t-j)(1+r)^{j-1}%=(s-1)^{-1}t( s^{t-1}-1)-(s-1)^{-1}ts^{t-1}+(s-1)^{-2}(s^t-1)
=(1+r)^{t}\sum_{j=2}^{t}(j-1)(1+r)^{-j}=r^{-2}((1+r)^t-rt-1).\]
On the other hand, we have
\begin{align*}
z_t^{(3)}-z_t^{(4)}&=(r_1r_3-r_2r_4)(t-1)+(1-r)(z_{t-1}^{(3)}-z_{t-1}^{(4)})\\
&=(r_1r_3-r_2r_4)\sum_{j=1}^{t-1}(t-j)(1-r)^{j-1}%=r_1(r_2- r_3)r_4^{-2}((1-r_4)^t+r_4t-1)
=2\alpha_4((1-r)^t+rt-1),
\end{align*}
so that
\begin{align*}
z_t^{(3)}%&=\tfrac{1}{2}r_1(r_2+ r_3)r_4^{-2}((1+r_4)^t-r_4t-1)+\tfrac{1}{2}r_1(r_2- r_3)r_4^{-2}((1-r_4)^t+r_4t-1)\\
 %&=\tfrac{1}{2}r_1(r_2+ r_3)r_4^{-2}(1+r_4)^t-r_1r_3r_4^{-1}t-r_1r_2r_4^{-2}+\tfrac{1}{2}r_1(r_2- r_3)r_4^{-2}(1-r_4)^t\\
 &=\alpha(1+r)^t-r_2r_4r^{-1}t-r_1r_3r^{-2}+\alpha_4(1-r)^t,\\
 z_{t}^{(4)}%&=\tfrac{1}{2}r_1(r_2+ r_3)r_4^{-2}((1+r_4)^t-r_4t-1)-\tfrac{1}{2}r_1(r_2- r_3)r_4^{-2}((1-r_4)^t+r_4t-1)\\
% &=\tfrac{1}{2}r_1(r_2+ r_3)r_4^{-2}(1+r_4)^t-r_1r_2r_4^{-1}t-r_1r_3r_4^{-2}-\tfrac{1}{2}r_1(r_2- r_3)r_4^{-2}(1-r_4)^t.
 &=\alpha(1+r)^t-r_1r_3r^{-1}t-r_2r_4r^{-2}-\alpha_4(1-r)^t.
\end{align*}

By   \eqref{EGO}, we have
$$c_t=z_t^{(2)}+z_t^{(3)}-x_t(0),$$
and the stated formula \eqref{Ukr} follows from the obtained expression for $z_t^{(3)}$ 
and the next consequence of \eqref{xt}:
 \begin{align}\label{xt0}
x_t(0)=r_2r_4^{-1}(1-(1-r_4)^{t}).
\end{align}

\vspace{5mm}
\noindent{\sc Proof of \eqref{Zel}}. 
Relation \eqref{yt0} implies
 \begin{align*}
y_t(1) %&=r_2z_{t-1}^{(1)}+r_4z_{t-1}^{(4)}+y_{t-1}(1)(1-r_4)\\
%&=r_1r_2(t-1)+\tfrac{1}{2}r_1(r_2+ r_3)r_4^{-1}(1+r_4)^{t-1}-r_1r_2(t-1)-r_1r_3r_4^{-1}-\tfrac{1}{2}r_1(r_2- r_3)r_4^{-1}(1-r_4)^{t-1}+y_{t-1}(1)(1-r_4)\\
&=\alpha r(1+r)^{t-1}-r_2r_4r^{-1}-\alpha_4r(1-r)^{t-1}+y_{t-1}(1)(1-r),
\end{align*}
which entails
 \begin{align*}
y_t(1)&=\sum_{j=1}^{t-1}\Bigg(\alpha r(1+r)^{t-j}-r_2r_4r^{-1}-\alpha_4r(1-r)^{t-j}\Bigg)(1-r)^{j-1}\\
&=\tfrac{1}{2}\alpha(1+r)((1+r)^{t-1}-(1-r)^{t-1})-r_2r_4r^{-2}(1-(1-r)^{t-1})-\alpha_4r(1-r)^{t-1}(t-1)\\
&=\tfrac{1}{2}\alpha(1+r)^{t}-r_2r_4r^{-2}-\alpha_4r(1-r)^{t-1}(t-1)+\alpha_5(1-r)^{t-1},
\end{align*}
where 
$\alpha_5=r_2r_4r^{-2}-\tfrac{1}{2}\alpha(1+r).$
Thus, 
$$y_t(1)\sim \tfrac{1}{2}\alpha(1+r)^{t},\quad t\to\infty.$$
Using this as the initiation  step for the induction over $m$ based on recursion \eqref{ytm}, we find  that
\begin{align*}
y_t(m)\sim 2^{-m}\alpha(1+r)^{t},\quad m\ge1,\quad t\to\infty.
\end{align*}
This gives  \eqref{Zel}  in view of \eqref{Ukr} and
$$c_t(m)=x_t(m)+y_t(m), \quad 1\le m\le t-1,$$
where $x_t(m)=o((1+r)^{t})$ in accordance with \eqref{xtm} and \eqref{xt0}.
%Variance $\sigma^2_t=\rV(Z_t)$

\begin{figure}[H]
\centering
\includegraphics[width=7cm]{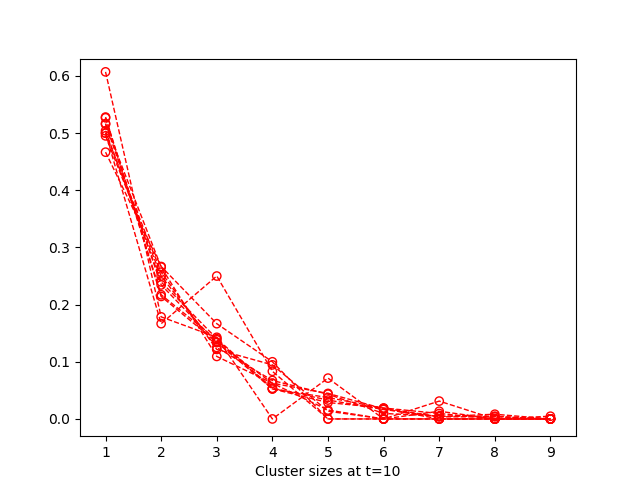} \includegraphics[width=7cm]{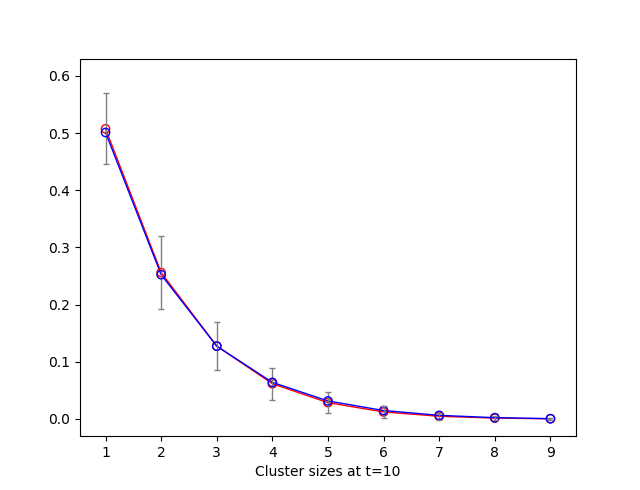}
 \caption{Simulation results for the proportions $C_{t}(m)/C_{t}$, $m=1,\ldots, t-1$ with $t=10$ and the amplification rates $(r_1,r_2,r_3,r_4,r)=(0.3,0.3,0.4,0.5,0.9)$. 
 The left panel presents ten individual simulation results. The right panel summarises 1 000 000 simulation results for the proportions $C_{t}(m)/C_{t}$: the red line connects the simulation averages,  the vertical intervals attached to the red line show the means 
 $\pm$ one standard deviations obtained from the simulations. Also on the right panel, the blue line connects the theoretical values for $\rE(C_{t}(m))/\rE(C_{t})$, this is the same blue line as on the right panel of Figure  \ref{f5}.}
 \label{M1}
\end{figure}

\section{Discussion}\label{Dis}

The number of approaches and applications that use UMIs in sequencing is rapidly increasing. In cancer
diagnostics, the use of UMIs is crucial since it allows to correct for both polymerase induced errors and
amplification biases \cite{JoG}.  Many sample types and matrices are challenging to analyse due to limited amounts of DNA and enzymatic inhibitors.
 In this paper we propose a convenient bookkeeping system for annotating the emerging UMI clusters during $t$ consecutive barcoding PCR cycles. 

The proposed tree based bookkeeping system leads to a branching process model for the counts $C_t(m)$ of the UMI clusters of sizes $m=0,1,\ldots,m-1$. Our model distinguishes between five PCR amplification rates $(r_1,r_2,r_3,r_4,r)$. A key feature of interest for such a model is the set of proportions $C_{t}(m)/C_{t}$, $m=1,\ldots, t-1$, where $C_t=C_{t}(1)+\ldots+C_{t}(t-1)$ is the total number of the UMI clusters.
 
The main theoretical finding of this paper, convergence  \eqref{Zel}, claims that the ratio between the expected counts
$\rE(C_t(m))/\rE(C_t)$ is approximately $2^{-m}$  regardless of the underlying parameters $(r_1,r_2,r_3,r_4,r)$. It was demonstrated that this approximation formula works well even for moderately large values of $t$, see  the right panel of Figure \ref{f5}. We \textit{hypothesise} a biologically more relevant asymptotic result
\begin{equation}\label{zela}
\rE(C_{t}(m)/C_t)\to 2^{-m},\quad m\ge 1,\quad t\to\infty.
\end{equation}
To address this hypothesis, a simulation study based on our model was performed by Hongui Zhan and Yizhe Gu, two master students at the Chalmers University of Technology. Their simulation results summarised by Figure \ref{M1}, support the approximation formula \eqref{zela} for the moderate value of $t=10$ and a particular choice of the amplification rates $(r_1,r_2,r_3,r_4,r)=(0.3,0.3,0.4,0.5,0.9)$.

%To overcome these issues as well as problems related library construction associated to UMIs the number of barcoding PCR cycles $t$ can be increased. However, when $t$ increases and r < 100\% it will affect our ability to quantify molecules accurately, since the relationship between number of UMIs compared to number of original molecules becomes complex. Here, we …. [Serik fyll på med info]

Our model of the barcoding PCR step uses five different amplification rates as the key model parameters $(r_1,r_2,r_3,r_4,r)$. 
  In sequencing, amplification rates are rarely assessed and  there is no general method to determine $(r_1,r_2,r_3,r_4)$. 
%In the framework of quantitative PCR, the overall amplification rate $r$ can be assessed by the standard curves \cite{SvD}. It is a well-established fact that $r$ varies between target molecules due to sequence context as well as between samples due to sample inhibition \cite{Ruj,Bar}. The amplification rate also decreases during the last PCR cycles when reagents become sparse. Scientist working with PCR and sequencing are experimentally used to fact that some samples and sequences suffer from poor amplification rates or that the original DNA molecules never become amplified. The early amplification rates when the first DNA molecules are amplified are challenging to determine by experimental approaches. 
%
In the framework of quantitative PCR, the overall amplification rate $r$  can be assessed by
standard curves \cite{SvD}.  The amplification rate $r$  varies between assays due to different sequence
context as well as between samples due to sample inhibition \cite{Bar, Ruj}. The amplification rate also decreases
during the last PCR cycles, when reagents become sparse. Scientist working with PCR and sequencing are
experimentally used to the fact that some samples and sequences suffer from poor amplification rates or that the
original DNA molecules never become amplified. 
In future studies, it will be important to verify our model with experimental data to estimate the importance of different model parameters. Such a verified model %that enables us to estimate the importance of different experimental parameters 
will be valuable in development of improved sequencing protocols and our ability to detect and quantify individual DNA molecules with single nucleotide resolution.   

%A preliminary analysis of experimental data obtained from the full protocol  involving two PCR steps (Figure  \ref{f02}) was recently performed in \cite{GZ}. This work produced promising results of estimating the amplification rates which are specific to the target molecules and the numbers of PCR cycles $t,x$ used in the first and the second PCR steps.
%Our forthcoming studies will aim at developing data analytical procedures to verify our model with experimental data and to estimate the different PCR amplification rates. A verified model that enable us to estimate the importance of different experimental parameters will be valuable in development of improved sequencing protocols and our ability to detect and quantify individual DNA molecules with single nucleotide resolution.   
% 

\subsection*{Acknowledgements} We are grateful to professor Peter Jagers for bringing us together and for fruitful discussions. 
This research was partially funded by Region Västra Götaland, Sweden; Swedish Cancer Society (20-1098); Swedish Research Council (2020-01008); Swedish Childhood Cancer Foundation (MTI2019-0008 and 2020-0007); the Swedish state under the agreement between the Swedish government and the county councils, the ALF-agreement (ALFGBG-965065); Sweden’s Innovation Agency and the Sjöberg Foundation.
%https://www.pnas.org/content/108/22/9026

%https://www.nature.com/articles/srep14629

\end{document}